\documentclass[a4paper]{article}

\usepackage{amsmath}
\usepackage{amssymb}
\usepackage[english]{babel}
\usepackage[small,bf]{caption}
\usepackage{cite}
\usepackage[T1]{fontenc}
\usepackage{graphicx}
\usepackage[hidelinks]{hyperref}
\usepackage{IEEEtrantools}
\usepackage[utf8]{inputenc}

\title{Enhancing Boolean networks with continuous logical operators and edge tuning}
\author{Arnaud Poret\textsuperscript{1}, Claudio Monteiro Sousa\textsuperscript{2}, Jean-Pierre Boissel\textsuperscript{3}}
\date{\today}

\begin{document}

\maketitle

\bigskip

\begin{abstract}
Due to the scarcity of quantitative details about biological phenomena, quantitative modeling in systems biology can be compromised, especially at the subcellular scale. One way to get around this is qualitative modeling because it requires few to no quantitative information. One of the most popular qualitative modeling approaches is the Boolean network formalism. However, Boolean models allow variables to take only two values, which can be too simplistic in some cases. The present work proposes a modeling approach derived from Boolean networks where continuous logical operators are used and where edges can be tuned. Using continuous logical operators allows variables to be more finely valued while remaining qualitative. To consider that some biological interactions can be slower or weaker than other ones, edge states are also computed in order to modulate in speed and strength the signal they convey. The proposed formalism is illustrated on a toy network coming from the epidermal growth factor receptor signaling pathway. The obtained simulations show that continuous results are produced, thus allowing finer analysis. The simulations also show that modulating the signal conveyed by the edges allows to incorporate knowledge about the interactions they model. The goal is to provide enhancements in the ability of qualitative models to simulate the dynamics of biological networks while limiting the need of quantitative information.
\end{abstract}

\bigskip

\noindent Copyright 2014-2019 Arnaud Poret, Claudio Monteiro Sousa, Jean-Pierre Boissel

\bigskip

\noindent This article is licensed under the Creative Commons Attribution-NonCommercial-NoDerivatives 4.0 International License. To view a copy of this license, visit \url{https://creativecommons.org/licenses/by-nc-nd/4.0/}.

\bigskip

\noindent \textsuperscript{1} \texttt{arnaud.poret@gmail.com} (corresponding author)\\
UMR CNRS 5558 Laboratoire de Biométrie et Biologie Évolutive, Lyon, France

\noindent \textsuperscript{2} \texttt{claudio.monteiro@novadiscovery.com}\\
Novadiscovery, Lyon, France

\noindent \textsuperscript{3} \texttt{jean-pierre.boissel@novadiscovery.com}\\
Novadiscovery, Lyon, France

\newpage

\tableofcontents

\section{Introduction}
Quantitative modeling in systems biology allows scientists to produce formal models of biological systems and then to implement them on computers \cite{cheong2008wires,mogilner2006quantitative}. With such computational models, scientists can perform \textit{in silico} experiments which have the advantage of being less costly in time and resources than the traditional wet-lab experiments. However, the stumbling block of \textit{in silico} approaches is that they are built from the available knowledge: not all is known about everything. Nevertheless, an impressive and ever increasing amount of biological knowledge is already available in the scientific literature, databases and knowledge bases such as KEGG \cite{tanabe2012kegg} and Reactome \cite{croft2014reactome}.

In addition to the difficulty of integrating an increasing body of knowledge comes the inherent complexity of biological systems themselves \cite{mazzocchi2012complexity,koch2012complexity,longtin2005integrated,bonchev2003complexity,oltvai2002complexity,weng1999complexity}: this is where computational tools can help owing to their integrative power \cite{altaf2014systems,dada2011multi,ghosh2011software,navlakha2011algorithms,sobie2011systems,greene2010integrative,boissel2009modeling,boissel2008modelling}. This interplay between wet-lab and computational biology is synergistic rather than competitive \cite{di2006vivo}. Wet-lab experiments produce factual results so that they can be considered as trustworthy sources of knowledge. Once these factual pieces of knowledge are obtained, computational tools can help to integrate them and infer new ones. This computationally obtained knowledge can subsequently be used to direct further wet-lab experiments, thus mutually potentiating the whole.

One of the main difficulties encountered when quantitatively modeling biological systems with, for example, systems of differential equations \cite{ilea2011ordinary} is that the required quantitative parameter values are often difficult to obtain due to experimental limitations, particularly at the subcellular scale. One solution to overcome this barrier is qualitative modeling because it requires few to no quantitative information while producing informative predictions \cite{wynn2012logic}.

Several qualitative modeling approaches already exist and are mostly based on logic \cite{ballerstein2013discrete,chaouiya2013logical,morris2010logic,watterson2008logic}, such as Boolean networks \cite{saadatpour2012boolean,albert2008boolean} which are based on Boolean logic \cite{boole1847mathematical}. However, this is at the cost of being qualitative: no quantification is performed. This does not mean that qualitative modeling is a downgrade of the quantitative one. This means that scientists have different approaches at their disposal, each with its advantages and disadvantages, depending on the pursued goals and the available resources. If accurate numerical results are expected, quantitative modeling is required. However, if tendencies and global properties are the main concern, qualitative modeling is entirely fitting and proved itself through several works \cite{von2014boolean,fumia2013boolean,grieco2013integrative,moreno2013modeling,peterson2013qualitative,tokar2013boolean,akman2012digital,rodriguez2012boolean,singh2012boolean,bhardwaj2011exploring,kang2011genetic,bauer2010receptor,naldi2010diversity,thakar2010boolean,ge2009boolean,mai2009boolean,sahin2009modeling,samaga2009logic,schlatter2009off,davidich2008boolean,kervizic2008dynamical,zhang2008network,gupta2007boolean,faure2006dynamical,mendoza2006network,huang2000shape}.

The present work proposes a continuous logic-based modeling approach aimed at enhancing the Boolean network formalism. The basic principles remain the same as in Boolean networks: given a biological network \cite{zhu2007getting,barabasi2004network,alm2003networks}, entities are modeled by variables and their interactions by functions allowing their value to be updated at each iteration of the simulation. However, Boolean operators are replaced by the operators of fuzzy logic \cite{bronshtein2007fuzzylogic,zadeh1988fuzzy}, allowing variables to be valued at any real number between $0$ and $1$. All the possible truth degrees between the absolutely true and the definitively false can therefore be considered.

The results obtainable with continuous logical operators, such as those of fuzzy logic, can be finer than those obtainable with Boolean operators. In some cases, the ON\slash OFF nature of Boolean logic is a relevant choice, as with gene regulatory networks where gene expression levels can be approximated by Boolean states \cite{emmert2014gene,xiao2009tutorial,karlebach2008modelling,rockett2006gene}. However, in some other cases where things are not necessarily binary, such as in signaling pathways where enzymes can be more or less active, using continuous logical operators can be an interesting choice.

In addition of using continuous logical operators, some additional features are introduced in order to capture more behavioral aspects of biological networks. These additional features concern the edges of the network. In the present work, the edges are seen as conveyors of signals corresponding to the influences exerted by entities of the network onto other ones. This signal, together with its modulation, are taken into account so that edges can be tuned. To do so, edge states are computed and the signal they convey can be slowed or weakened. This results in a qualitative modeling approach intended to bring a fine qualitative quantification of biological network behaviors.

Talking about a qualitative quantification can appear somewhat contradictory but is common in thinking processes, which are at the basis of any scientific reasoning. Simple examples of such qualitative quantifications could be to state that an enzyme is more active than another one, or to state that an enzyme is moderately active: quantification is expressed by perceptions and tendencies. Indeed, qualitative quantifications are expressed by words rather than measurements, hence their qualitative nature, and are characteristic of fuzzy logic \cite{zadeh2001computing,zadeh1996fuzzy}.

It should be noted that fuzzy logic-based modeling is a promising approach successfully developed in several works \cite{cleophas2014fuzzy,morris2011training,sun2010new,aldridge2009fuzzy,huang2009fuzzy,franco2007application,du2005modeling,lee1999incorporating,wang1996complex}. However, the present work is not fuzzy logic-based: there are no fuzzy sets, no fuzzy membership functions, no degrees of membership and no fuzzy inference systems. Only the logical operators are taken from fuzzy logic to replace the Boolean ones, the goal being to enhance the Boolean network formalism by extending it to a continuous formalism, and by adding edge tuning.

\section{Methods}
This section recalls some basic principles, namely biological and Boolean networks. It also introduces the continuous logical operators taken from fuzzy logic and then describes how the proposed enhancement of Boolean networks is built. An example together with its implementation are also described.

\subsection{Basic principles}

\subsubsection{Biological networks}

A biological network is a way to conceptualize a set of interacting entities where entities are represented by nodes and interactions by edges. It is based on graph theory \cite{ma2009insights,ma2008network,bronshtein2007graphtheory,huber2007graphs,mason2007graph,aittokallio2006graph}, thus bringing formal tools to encode information about biological systems, particularly their topology. Moreover, biological networks offer a convenient visualization \cite{larkin1987diagram} of the complex interconnections lying in biological systems. As said Napoleon Bonaparte:
\begin{quote}
``A good sketch is better than a long speech.''
\end{quote}

Several types of biological networks can be encountered depending on the scale, the involved entities and their interconnections. For example, at the ecological scale, food webs are biological networks where nodes represent species and edges represent trophic relations \cite{thompson2012food,ings2009review,schmitz2007food}. At the subcellular scale there is, for example, gene regulatory networks where nodes represent gene products and edges represent gene expression regulations. Whatever is the scale or the entities, the principles remain the same: given a biological system, nodes represent entities and edges represent interactions between them.

Mathematically, a network is a digraph $G=(V,E)$ where $V=\lbrace v_{1},\ldots,v_{n}\rbrace$ is the set containing the nodes of the network and $E=\lbrace (v_{i,1},v_{j,1}),\ldots,(v_{i,m},v_{j,m})\rbrace$ is the set containing the edges linking these nodes. In practice, nodes represent entities while edges represent binary relations $R \subset V^{2}$ involving them: $v_{i}\ R\ v_{j}$ \cite{zhu2007getting}. It indicates that the node $v_{i}$ exerts an influence on the node $v_{j}$. For example, in gene regulatory networks \cite{emmert2014gene}, $v_{i}$ can be a transcription factor while $v_{j}$ another gene product. Edges are frequently signed so that they indicate if $v_{i}$ exerts a positive or a negative influence on $v_{j}$, such as an activation or an inhibition.

\subsubsection{Boolean networks}

Boolean networks, pioneered in biology by Kauffman \cite{kauffman1969metabolic}, Ostrander \cite{ostrander1973functional}, Thomas \cite{thomas1973boolean} and Glass \cite{glass1975classification}, are one of the existing qualitative modeling approaches. While being conceptually simple, Boolean networks are able to predict and reproduce features of biological systems and then to bring relevant insights \cite{albert2014boolean,wang2012boolean,bornholdt2008boolean,huang2001genomics,leclerco1983boolean}. This makes them an attractive and efficient approach, especially when the complexity of biological systems renders quantitative approaches unfeasible due to the amount of quantitative details they require.

As their name indicates, Boolean networks are based on Boolean logic and, like biological networks, are also based on graph theory: nodes represent Boolean variables and edges represent interdependencies between them. Boolean networks can be classified according to their updating scheme as synchronous or asynchronous: if all the variables are updated simultaneously at each iteration of the simulation then the network is synchronous, otherwise it is asynchronous. While there is one synchronous updating scheme, several asynchronous updating schemes exist:
\begin{itemize}
\item the random order asynchronous updating scheme where, at each iteration, an updating order is randomly selected for the variables
\item the general asynchronous updating scheme where, at each iteration, a randomly selected variable is updated
\item the deterministic asynchronous updating scheme where a divisor is assigned to each variable and then, at each iteration, a variable is updated if and only if the iteration is a multiple of its divisor
\end{itemize}

With the exception of deterministic asynchronous Boolean networks, only synchronous Boolean networks are deterministic because at each iteration the variables have only one possible successor. This makes synchronous Boolean networks easier to compute than asynchronous ones \cite{garg2008synchronous}. However, when the dynamics of a biological network is computed synchronously, it is assumed that all its components evolve simultaneously, an assumption which can be inappropriate according to what is modeled.

Mathematically, a Boolean network is a network where nodes are Boolean variables $x_{i}$ and edges $(x_{i},x_{j})$ are the $input\_of$ relation: $x_{i}\ input\_of\ x_{j}$. Each variable $x_{i}$ has $b_{i} \in [\![0,n]\!]$ inputs influencing its state. Note that $b_{i}=0$ is possible. In this case, $x_{i}$ is an input of the network. Depending on the updating scheme, at each iteration $k \in [\![k_{0},k_{end}]\!]$ one or more variables $x_{i}$ are updated using their associated Boolean updating function $f_{i}$. This function uses Boolean operators, typically $\land$ ($and$), $\lor$ ($or$) and $\lnot$ ($not$), to specify how the inputs $x_{i,1},\ldots,x_{i,b_{i}}$ of $x_{i}$ have to be related to compute its value, as in the following pseudocode representing a synchronous updating:\\

\noindent \textbf{for} $k \gets k_{0},\ldots,k_{end}$\\
\indent $x_{1} \gets f_{1}(x_{1,1},\ldots,x_{1,b_{1}})$\\
\indent $\vdots$\\
\indent $x_{n} \gets f_{n}(x_{n,1},\ldots,x_{n,b_{n}})$\\
\textbf{end for}\\

\noindent which can be written in a more concise form:\\

\noindent \textbf{for} $k \gets k_{0},\ldots,k_{end}$\\
\indent $\boldsymbol{x} \gets \boldsymbol{f}(\boldsymbol{x})$\\
\textbf{end for}\\

\noindent where $\boldsymbol{f}=(f_{1},\ldots,f_{n})$ is the Boolean updating function of the network and $\boldsymbol{x}=(x_{1},\ldots,x_{n})$ is its state vector. The value of the state vector belongs to the state space $S=\lbrace 0,1\rbrace^{n}$ which is the set containing all the possible states of the network.

\subsubsection{Fuzzy operators}

The main difference between Boolean and fuzzy logic is that the former is discrete, that is valued in $[\![0;1]\!] \subset \mathbb{N}$, whereas the latter can be continuous, that is valued in $[0;1] \subset \mathbb{R}$. Fuzzy logic can be seen as a generalization of the Boolean one, implying that the fuzzy counterparts of the Boolean operators behave like them on $[\![0;1]\!]$ while being defined on $[0;1]$. The generalization of the Boolean $AND$ operator is the $t\text{-}norm$, the generalization of the Boolean $OR$ operator is the $s\text{-}norm$ and the generalization of the Boolean $NOT$ operator is the $complement$:

\begin{small}
\begin{IEEEeqnarray*}{r l C l r C l}
t\text{-}norm\colon&[0;1]^{2}&\to&[0;1]\colon&(x,y)&\mapsto&t\text{-}norm(x,y)\\
s\text{-}norm\colon&[0;1]^{2}&\to&[0;1]\colon&(x,y)&\mapsto&s\text{-}norm(x,y)\\
complement\colon&[0;1]&\to&[0;1]\colon&x&\mapsto&complement(x)
\end{IEEEeqnarray*}
\end{small}

\noindent where $x,y \in [0;1]$. There exist several mathematical formulations of the $t\text{-}norm$, $s\text{-}norm$ and $complement$, all fulfilling the rules of Boolean algebra \cite{bronshtein2007booleanalgebra} but defined on $[0;1]$. For convenience, both the Boolean and fuzzy operators can be named $AND$, $OR$ and $NOT$, the context specifying which of them is referred to.

Due to the ability of fuzzy operators to be continuous, the variables can take their value in $[0;1]$. Therefore, they can be equal to $1$ (true), $0$ (false) or all the other real numbers of $[0;1]$ (more or less true): all the truth degrees between true and false are considered. This can be more realistic in a world where things are not necessarily binary. For example, a Boolean model of a signaling pathway allows enzymes to be ON or OFF but nothing between. However, one can expect an enzyme to be in an intermediate activity level, an expectation easily implementable with continuous logic-based models. Whatever the truth degrees represent, using continuous logical operators enables to consider all the intermediate levels of what is modeled without leaving the qualitative modeling formalism.

\subsection{The proposed logic-based modeling}

First of all, it should be mentioned that a distinction is made between quantitative and qualitative parameters, this distinction residing in what parameters translate. A quantitative parameter translates a quantification obtained by experimental measurements whereas a qualitative parameter translates a perception by means of truth degrees. For example, regarding the velocity of a biochemical reaction, ``slow'' could be expressed by the truth degree $0.2$ whereas ``fast'' by $0.8$: this is the truth degree of the statement ``This biochemical reaction is fast.''. Unlike experimental quantifications which are \textit{de facto} objective, perceptions are subjective, so the same applies to their associated truth degrees. Incorporating qualitative parameters should not yield the scarcity of parameter values encountered in quantitative modeling because qualitative information is easier to obtain.

To build the proposed logic-based modeling from Boolean networks, the Boolean operators $AND$, $OR$ and $NOT$ are replaced by the fuzzy operators $t\text{-}norm$, $s\text{-}norm$ and $complement$. The initial states $x_{i}(k_{0})$ of the variables $x_{i}$ now belong to the interval of real numbers $[0;1]$. Consequently, the variables evolve in $[0;1]$ and their associated updating functions $f_{i}$ become functions from $[0;1]^{n}$ to $[0;1]$:

\begin{small}
\begin{IEEEeqnarray*}{c}
f_{i}\colon [0;1]^{n} \to [0;1]\colon \boldsymbol{x} \mapsto f_{i}(\boldsymbol{x})
\end{IEEEeqnarray*}
\end{small}

\noindent Corollary, the state vector $\boldsymbol{x}$ belongs to $[0;1]^{n}$ and its updating function $\boldsymbol{f}$ becomes a function from $[0;1]^{n}$ onto itself:

\begin{small}
\begin{IEEEeqnarray*}{c}
\boldsymbol{f}\colon [0;1]^{n} \to [0;1]^{n}\colon \boldsymbol{x} \mapsto \boldsymbol{f}(\boldsymbol{x})
\end{IEEEeqnarray*}
\end{small}

Some additional features are added in order to capture more behavioral aspects of biological networks. These features concern the edges of the network and are presented separately for the sake of clarity before being integrated all together.

\subsubsection{Edge computation}

As with node states, edge states are computed. For convenience, edges are notated $e_{ij}$ instead of $(x_{i},x_{j})$. An edge $e_{ij}$ is seen as a channel conveying the signal sent by its source node $x_{i}$ to its target node $x_{j}$. Practically, $e_{ij}$ conveys the value $x_{i}(k)$ to $x_{j}$ and then $f_{j}$ uses it to compute $x_{j}(k+1)$. This is implicitly done in Boolean networks where $x_{j}(k+1)=f_{j}(\ldots,x_{i}(k),\ldots)$. In the present work this is made explicit in order to modulate the signal conveyed by the edges. Consequently, the updating functions $f_{j}$ no longer directly accept the $x_{i}(k)$ as arguments but accept the $e_{ij}(k)$.

Since $e_{ij}$ conveys $x_{i}(k)$, its value $e_{ij}(k)$ should be $x_{i}(k)$, but this is where additional features are added. Indeed, an updating function $f^{edge}_{ij}$ is attributed to each edge $e_{ij}$:

\begin{small}
\begin{IEEEeqnarray*}{c}
e_{ij}(k+1)=f^{edge}_{ij}(x_{i}(k),e_{ij}(k))
\end{IEEEeqnarray*}
\end{small}

\noindent Note that in addition to the value $x_{i}(k)$ of the source node $x_{i}$, $f^{edge}_{ij}$ also takes as argument the value $e_{ij}(k)$ of the edge $e_{ij}$ itself. This is required for the additional feature \textit{edge reactivity} described below.

As mentioned above, the updating functions $f_{j}$ now accept the $e_{ij}(k)$ instead of the $x_{i}(k)$. For convenience, the updating functions $f_{j}$ of the nodes are renamed $f^{node}_{j}$:

\begin{small}
\begin{IEEEeqnarray*}{c}
x_{j}(k+1)=f^{node}_{j}(\boldsymbol{e}(k))
\end{IEEEeqnarray*}
\end{small}

\noindent where $\boldsymbol{e}=(\ldots,e_{ij},\ldots)$ is the counterpart of $\boldsymbol{x}=(\ldots,x_{i},\ldots)$, namely the state vector of the edges. Its value at the iteration $k$ is $\boldsymbol{e}(k)=(\ldots,e_{ij}(k),\ldots)$. Therefore, the node updating function $\boldsymbol{f}$ of the network becomes $\boldsymbol{f}^{node}=(\ldots,{f}^{node}_{i},\ldots)$:

\begin{small}
\begin{IEEEeqnarray*}{c}
\boldsymbol{x}(k+1)=\boldsymbol{f}^{node}(\boldsymbol{e}(k))
\end{IEEEeqnarray*}
\end{small}

\noindent and its counterpart the edge updating function $\boldsymbol{f}^{edge}=(\ldots,f^{edge}_{ij},\ldots)$ of the network is introduced:

\begin{small}
\begin{IEEEeqnarray*}{c}
\boldsymbol{e}(k+1)=\boldsymbol{f}^{edge}(\boldsymbol{x}(k),\boldsymbol{e}(k))
\end{IEEEeqnarray*}
\end{small}

On the basis of the updating scheme of synchronous Boolean networks, the computation becomes:\\

\noindent \textbf{for} $k \gets k_{0},\ldots,k_{end}$\\
\indent $\vdots$\\
\indent $e_{ij}(k+1) \gets f^{edge}_{ij}(x_{i}(k),e_{ij}(k))$\\
\indent $\vdots$\\
\indent $x_{i}(k+1) \gets f^{node}_{i}(\ldots,e_{ij}(k),\ldots)$\\
\indent $\vdots$\\
\textbf{end for}\\

\noindent which can be written in a more concise form:\\

\noindent \textbf{for} $k \gets k_{0},\ldots,k_{end}$\\
\indent $\boldsymbol{e}(k+1) \gets \boldsymbol{f}^{edge}(\boldsymbol{x}(k),\boldsymbol{e}(k))$\\
\indent $\boldsymbol{x}(k+1) \gets \boldsymbol{f}^{node}(\boldsymbol{e}(k))$\\
\textbf{end for}

\subsubsection{Edge reactivity}

The additional feature \textit{edge reactivity} is implemented by a qualitative parameter $r_{ij} \in [0;1]$ attributed to each edge $e_{ij}$. $r_{ij}$ is the portion of the signal conveyed by $e_{ij}$ which is updated at each iteration $k$, namely the portion of $e_{ij}(k)$ which is updated to $x_{i}(k)$:

\begin{small}
\begin{IEEEeqnarray*}{c}
e_{ij}(k+1)=(1-r_{ij}) \cdot e_{ij}(k)+r_{ij} \cdot x_{i}(k)
\end{IEEEeqnarray*}
\end{small}

\noindent The higher $r_{ij}$ is, the higher is the portion of $e_{ij}(k)$ which is updated: a highly reactive edge has a $r_{ij}$ close to $1$ whereas a poorly reactive edge has a $r_{ij}$ close to $0$.

Biologically, \textit{edge reactivity} takes into account that some biological interactions can be slower (or of higher inertia) than other ones. For example, an edge modeling the activation of a gene product expression by a transcription factor should have a lower $r_{ij}$ than an edge modeling an activating phosphorylation of an enzyme by another one. Indeed, gene expression is a complex mechanism involving several steps and then, relative to a phosphorylation, takes more time to be accomplished.

It is important to note that \textit{edge reactivity} both impacts the appearance and disappearance of the signal conveyed by the edges. For example, an edge with an \textit{edge reactivity} close to $0$ has a high inertia: once the source node is activated, the signal appears slowly in the edge and, once the source node is deactivated, the signal disappears slowly. In other words, an edge with a low reactivity (a high inertia) slowly implements the changes emitted by its source node.

\subsubsection{Edge weakening}

The additional feature \textit{edge weakening} is implemented by a qualitative parameter $w_{ij} \in [0;1]$ attributed to each edge $e_{ij}$. $w_{ij}$ is a weakening coefficient applied at each iteration $k$ on the signal conveyed by $e_{ij}$, that is on $x_{i}(k)$:

\begin{small}
\begin{IEEEeqnarray*}{c}
e_{ij}(k+1)=w_{ij} \cdot x_{i}(k)
\end{IEEEeqnarray*}
\end{small}

\noindent The higher $w_{ij}$ is, the lower is the weakening of the signal $x_{i}(k)$ conveyed by $e_{ij}$: a strong edge has a $w_{ij}$ close to $1$ whereas a weak edge has a $w_{ij}$ close to $0$.

Biologically, \textit{edge weakening} takes into account that some biological interactions can be weaker than other ones. For example, given a membrane receptor, an edge modeling its activation by a partial agonist should have a lower $w_{ij}$ than an edge modeling its activation by a full agonist.

\subsubsection{Combining the all}

\textit{Edge reactivity} and \textit{edge weakening} are described separately for the sake of clarity but are both computed at each iteration:

\begin{small}
\begin{IEEEeqnarray*}{c}
e_{ij}(k+1)=(1-r_{ij}) \cdot e_{ij}(k)+r_{ij} \cdot w_{ij} \cdot x_{i}(k)
\end{IEEEeqnarray*}
\end{small}

\noindent hence the mathematical formulation of the edge updating functions $f^{edge}_{ij}$:

\begin{small}
\begin{IEEEeqnarray*}{c}
f^{edge}_{ij}(x_{i},e_{ij})=(1-r_{ij}) \cdot e_{ij}+r_{ij} \cdot w_{ij} \cdot x_{i}
\end{IEEEeqnarray*}
\end{small}

\subsection{Implementation}

In the present work, $k$ is not the time, it only represents the iterations performed during a run. Although quantifying time through $k$ is possible, here the goal is to visualize sequences of events linked by causal connections without time quantification. To do so, $k_{0}=1$ and $k_{end}=50$. The initial state $e_{ij}(k_{0})$ of each edge $e_{ij}$ is assumed to be equal to the initial state $x_{i}(k_{0})$ of its source node $x_{i}$: $e_{ij}(k_{0})=x_{i}(k_{0})$. To illustrate the proposed logic-based modeling approach, it is implemented on a toy network using R \cite{R}. The code is freely available on GitHub at \url{https://github.com/arnaudporet/smoosim}.

\subsubsection{Example}

\begin{figure*}[!hbp]
\begin{center}
\includegraphics[width=1\textwidth]{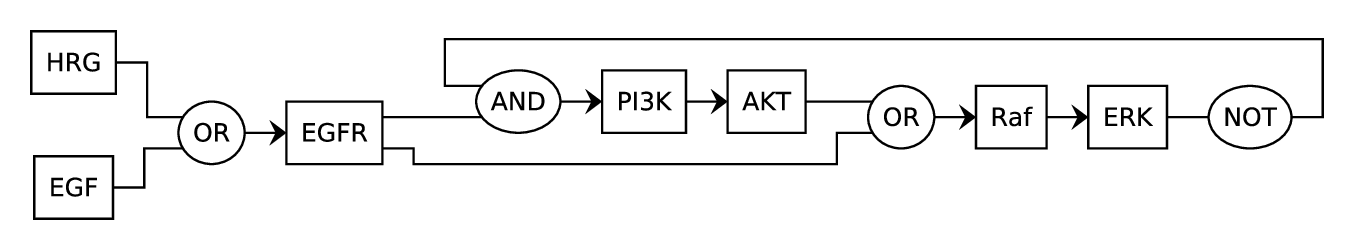}
\caption[Example]{\label{EGFR} Graphical representation of the example. Nodes are rectangles whereas logical gates are ellipses. This digraph should be read from left to right. For example, the node $PI3K$ is an input of the node $AKT$ whereas the node $ERK$, due to a feedback loop, is an input of the node $PI3K$. Logical gates are not nodes and, as such, edges only pass through them. For example, the edge $(ERK,PI3K)$ passes through a $NOT$ gate then an $AND$ gate whereas the edge $(Raf,ERK)$ does not pass through any logical gate.}
\end{center}
\end{figure*}

Although the presented logic-based modeling approach was used at Novadiscovery for the SysCLAD European project \cite{pison2014prediction} on a large biological network modeling the pathological mechanisms of chronic lung allograft dysfunction, here the example is a tiny sample of the epidermal growth factor receptor signaling pathway \cite{oda2005comprehensive} adapted from \cite{morris2010logic}. It is chosen for its simplicity so that it can be mentally computed in order to easily appreciate the results. The example is depicted in \hyperref[EGFR]{\texttt{Figure \ref*{EGFR}}} page \pageref{EGFR}. Below are the corresponding Boolean functions where $AND$, $NOT$ and $OR$ stand for the Boolean operators:

\begin{small}
\begin{IEEEeqnarray*}{r C l}
EGF(k+1)&=&\text{input set manually}\\
HRG(k+1)&=&\text{input set manually}\\
EGFR(k+1)&=&OR(EGF(k),HRG(k))\\
PI3K(k+1)&=&AND(EGFR(k),NOT(ERK(k)))\\
AKT(k+1)&=&PI3K(k)\\
Raf(k+1)&=&OR(EGFR(k),AKT(k))\\
ERK(k+1)&=&Raf(k)
\end{IEEEeqnarray*}
\end{small}

By applying the above-described methodology, below are the obtained updating functions $f^{edge}_{ij}$ and $f^{node}_{i}$ where $AND$, $NOT$ and $OR$ stand for the operators taken from fuzzy logic:

\begin{small}
\begin{IEEEeqnarray*}{r C l}
(EGF,EGFR)(k+1)&=&(1-r_{EGF,EGFR}) \cdot (EGF,EGFR)(k)\\
&&+r_{EGF,EGFR} \cdot w_{EGF,EGFR} \cdot EGF(k)\\
(HRG,EGFR)(k+1)&=&(1-r_{HRG,EGFR}) \cdot (HRG,EGFR)(k)\\
&&+r_{HRG,EGFR} \cdot w_{HRG,EGFR} \cdot HRG(k)\\
(EGFR,PI3K)(k+1)&=&(1-r_{EGFR,PI3K}) \cdot (EGFR,PI3K)(k)\\
&&+r_{EGFR,PI3K} \cdot w_{EGFR,PI3K} \cdot EGFR(k)\\
(ERK,PI3K)(k+1)&=&(1-r_{ERK,PI3K}) \cdot (ERK,PI3K)(k)\\
&&+r_{ERK,PI3K} \cdot w_{ERK,PI3K} \cdot ERK(k)\\
(PI3K,AKT)(k+1)&=&(1-r_{PI3K,AKT}) \cdot (PI3K,AKT)(k)\\
&&+r_{PI3K,AKT} \cdot w_{PI3K,AKT} \cdot PI3K(k)\\
(EGFR,Raf)(k+1)&=&(1-r_{EGFR,Raf}) \cdot (EGFR,Raf)(k)\\
&&+r_{EGFR,Raf} \cdot w_{EGFR,Raf} \cdot EGFR(k)\\
(AKT,Raf)(k+1)&=&(1-r_{AKT,Raf}) \cdot (AKT,Raf)(k)\\
&&+r_{AKT,Raf} \cdot w_{AKT,Raf} \cdot AKT(k)\\
(Raf,ERK)(k+1)&=&(1-r_{Raf,ERK}) \cdot (Raf,ERK)(k)\\
&&+r_{Raf,ERK} \cdot w_{Raf,ERK} \cdot Raf(k)
\end{IEEEeqnarray*}
\end{small}

\begin{small}
\begin{IEEEeqnarray*}{r C l}
EGF(k+1)&=&\text{input set manually}\\
HRG(k+1)&=&\text{input set manually}\\
EGFR(k+1)&=&OR((EGF,EGFR)(k),(HRG,EGFR)(k))\\
PI3K(k+1)&=&AND((EGFR,PI3K)(k),NOT((ERK,PI3K)(k)))\\
AKT(k+1)&=&(PI3K,AKT)(k)\\
Raf(k+1)&=&OR((EGFR,Raf)(k),(AKT,Raf)(k))\\
ERK(k+1)&=&(Raf,ERK)(k)
\end{IEEEeqnarray*}
\end{small}

\noindent It should be noted that $f^{node}_{EGF}$ and $f^{node}_{HRG}$ do not accept any $e_{ij}(k)$ as argument. This is because they are associated to the two inputs $EGF$ and $HRG$ of the network and are therefore set manually.

\subsubsection{Fuzzy operators}

As mentioned above, there exist several mathematical formulations of the logical operators of fuzzy logic, all fulfilling the rules of Boolean algebra but defined on the interval of real numbers $[0;1]$. In the present work, the algebraic formulation is used:

\begin{small}
\begin{IEEEeqnarray*}{r C l}
AND(x,y)&=&x \cdot y\\
OR(x,y)&=&x+y-x \cdot y\\
NOT(x)&=&1-x
\end{IEEEeqnarray*}
\end{small}

\noindent which is one of the most simple and convenient because, as its name indicates, consists in algebraic equations. Of course, one can choose another mathematical formulation of the fuzzy operators.

\subsubsection{Additional features}

Since $r_{ij} \in [0;1]$, its value can be set to any real number of $[0;1]$. However, $r_{ij}$ is a qualitative parameter. Rather than requiring it to be precisely valued as in quantitative modeling, its value is randomly picked in specified intervals of $[0;1]$ from a uniform distribution. By the way, this random selection introduces a little of a rudimentary stochasticity although introducing randomness is not the purpose of the present work. To do so, $[0;1]$ is split into intervals of truth degrees reflecting various subjective valuation of \textit{edge reactivity} expressed by words:
\begin{small}
\begin{center}
\begin{tabular}{l|l}
instantaneous&$r_{ij}=1$\\
faster&$r_{ij} \in [0.75;1]$\\
fast&$r_{ij} \in [0.5;0.75]$\\
slow&$r_{ij} \in [0.25;0.5]$\\
slower&$r_{ij} \in [0;0.25]$\\
down&$r_{ij}=0$
\end{tabular}
\end{center}
\end{small}
plus the entire interval $[0;1]$ in case of an undetermined \textit{edge reactivity}. For example, $r_{ij}=fast$ means that the value of $r_{ij}$ is randomly picked in $[0.5;0.75]$ from a uniform distribution.

This random selection occurs before each run and, once selected, the value of $r_{ij}$ remains the same during the run. To better approach the behavior of the modeled biological network, replicates are made: $p$ runs are performed and the results are superposed. Here, $p=5$. $w_{ij}$ and $x_{i}(k_{0})$ are subjected to the same replication with the following splits of $[0;1]$:
\begin{small}
\begin{center}
\begin{tabular}{l|l}
strong&$w_{ij}=1$\\
weak&$w_{ij} \in [0.75;1]$\\
weaker&$w_{ij} \in [0.5;0.75]$\\
faint&$w_{ij} \in [0.25;0.5]$\\
fainter&$w_{ij} \in [0;0.25]$\\
down&$w_{ij}=0$
\end{tabular}
\end{center}
\end{small}

\begin{small}
\begin{center}
\begin{tabular}{l|l}
full&$x_{i}(k_{0})=1$\\
much more&$x_{i}(k_{0}) \in [0.75;1]$\\
much&$x_{i}(k_{0}) \in [0.5;0.75]$\\
few&$x_{i}(k_{0}) \in [0.25;0.5]$\\
fewer&$x_{i}(k_{0}) \in [0;0.25]$\\
none&$x_{i}(k_{0})=0$
\end{tabular}
\end{center}
\end{small}
plus the entire interval $[0;1]$ in case of an undetermined \textit{edge weakening}\slash initial state. Of course, one can split $[0;1]$ in a different way. Moreover, one can imagine these qualitative parameters to be functions instead of constants.

\section{Results}
In this section, results obtained with the example through five different simulations are presented. Although the obtained curves are continuous due to the use of continuous logical operators, they are not quantitative. As qualitative results, rather than looking for numerical values, one can say for example that $PI3K$ is totally inhibited or that $ERK$ is partly activated, two simple examples of qualitative quantifications expressed by words and perceptions.

\subsection{Simulation 1}

$EGF$ and $HRG$ are the two inputs of the example and, since both can activate $EGFR$, one is sufficient to initiate the signaling cascade. It is assumed that at the resting state both the inputs are down: $\forall k$, $EGF(k)=HRG(k)=none$. However, at $k_{EGF}=k_{end}/10$, $EGF$ is activated: $\forall k \geq k_{EGF}$, $EGF(k+1)=full$. Therefore, $f^{node}_{EGF}$ and $f^{node}_{HRG}$ become:

\begin{small}
\begin{IEEEeqnarray*}{r C l}
EGF(k+1)&=&
\begin{cases}
full&\text{if }k \geq k_{EGF}\\
none&\text{if }k<k_{EGF}
\end{cases}\\
HRG(k+1)&=&none
\end{IEEEeqnarray*}
\end{small}

\begin{figure*}[!hbp]
\begin{center}
\includegraphics[width=1\textwidth]{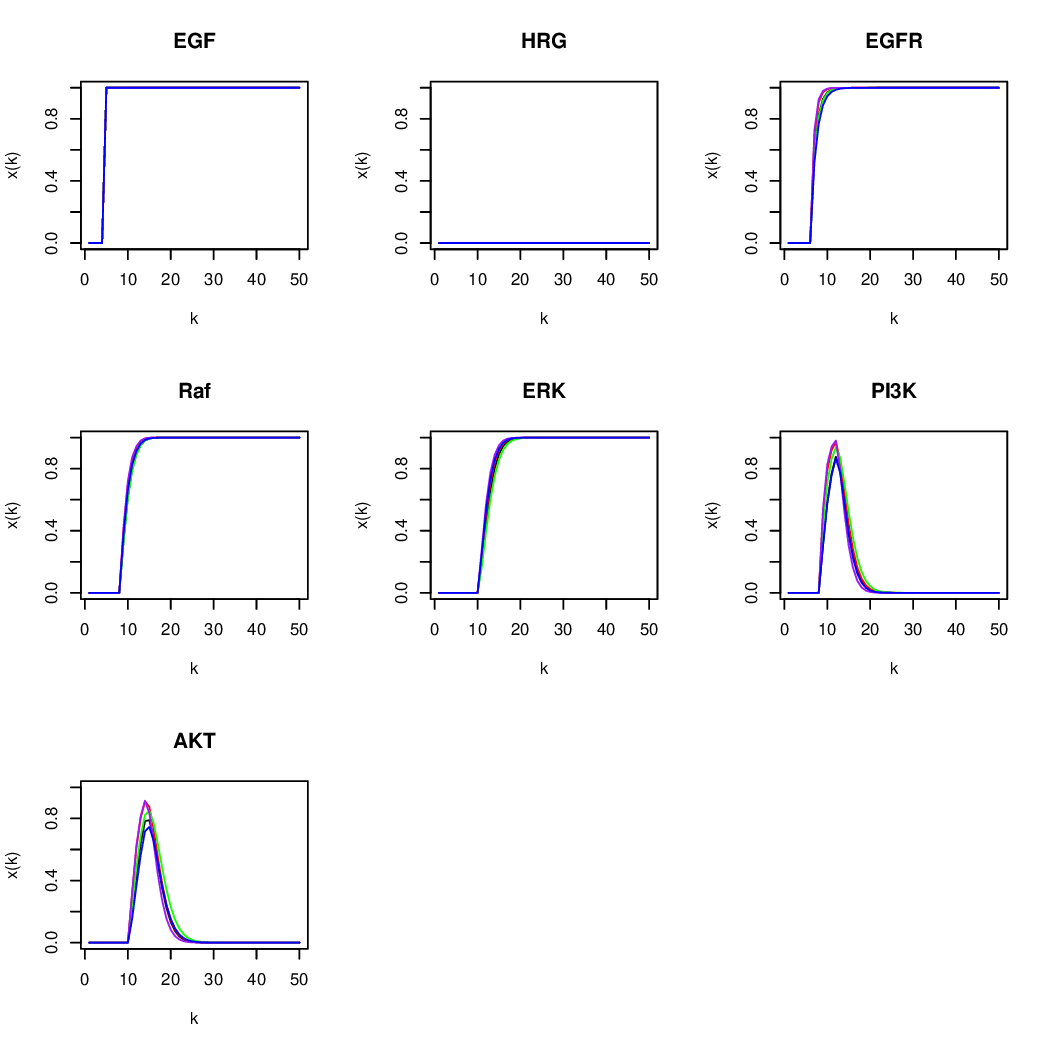}
\caption[Simulation 1]{\label{simulation1} Simulation 1: activation of $EGF$.}
\end{center}
\end{figure*}

\noindent The network being assumed to be at the resting state, $\boldsymbol{x}_{0}=(\ldots,none,\ldots)$. The $r_{ij}$ are set to $fast$ and the $w_{ij}$ are set to $strong$. The corresponding results are shown in \hyperref[simulation1]{\texttt{Figure \ref*{simulation1}}} page \pageref{simulation1}.

As expected, before $EGF$ activation the network is at rest: the signaling cascade is not active. However, once $EGF$ is activated the signaling cascade is activated too. This ultimately activates $ERK$, hence the subsequent inactivation of $PI3K$ despite sustained $EGFR$ activity. Since $AKT$ is activated by $PI3K$, it is also deactivated.

\subsection{Simulation 1 \textit{bis}}

\begin{figure*}[!hbp]
\begin{center}
\includegraphics[width=1\textwidth]{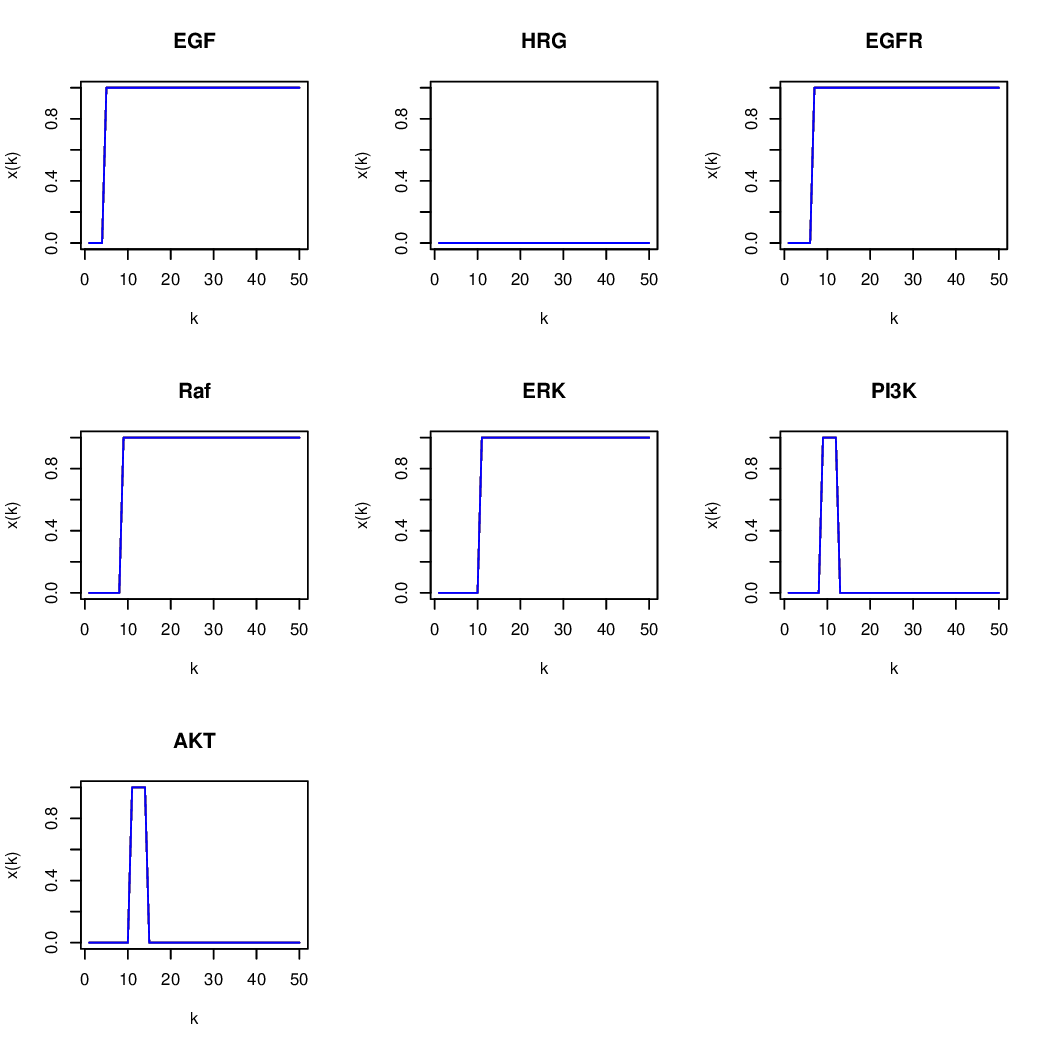}
\caption[Simulation 1 \textit{bis}]{\label{simulation1bis} Simulation 1 \textit{bis}: simulation 1 in the Boolean case.}
\end{center}
\end{figure*}

In order to compare the proposed logic-based modeling formalism with the classical Boolean one, the simulation 1 is reproduced in the Boolean case. Because the fuzzy operators also work with Boolean logic, the Boolean case can easily be obtained by setting the $r_{ij}$ to $instantaneous$ (full reactivity), the $w_{ij}$ to $strong$ (no weakening) and the $x_{i}(k_{0})$ to $none$ or $full$ ($0$ or $1$). Therefore, all is as in simulation 1 except the $r_{ij}$ which are set to $instantaneous$. The corresponding results are shown in \hyperref[simulation1bis]{\texttt{Figure \ref*{simulation1bis}}} page \pageref{simulation1bis}.

As expected, the Boolean case is obtained: the variables are binary and, \textit{de facto}, can not be continuous. Moreover, tuning the edges in term of reactivity and weakening is not possible. This shows that the Boolean case can capture only two nuances: all or nothing, unlike the continuous logic-base modeling formalism proposed in the present work.

\subsection{Simulation 2}

In addition to the inputs described in simulation 1, a perturbation is introduced. It consists in disabling the inhibitory effect of $ERK$ on $PI3K$, that is in disabling the edge $(ERK,PI3K)$. It points out an advantage of computing edge states: disturbing a node disturbs all its effects while selectively disturbing an edge prevents this, therefore allowing finer perturbations.

\begin{figure*}[!hbp]
\begin{center}
\includegraphics[width=1\textwidth]{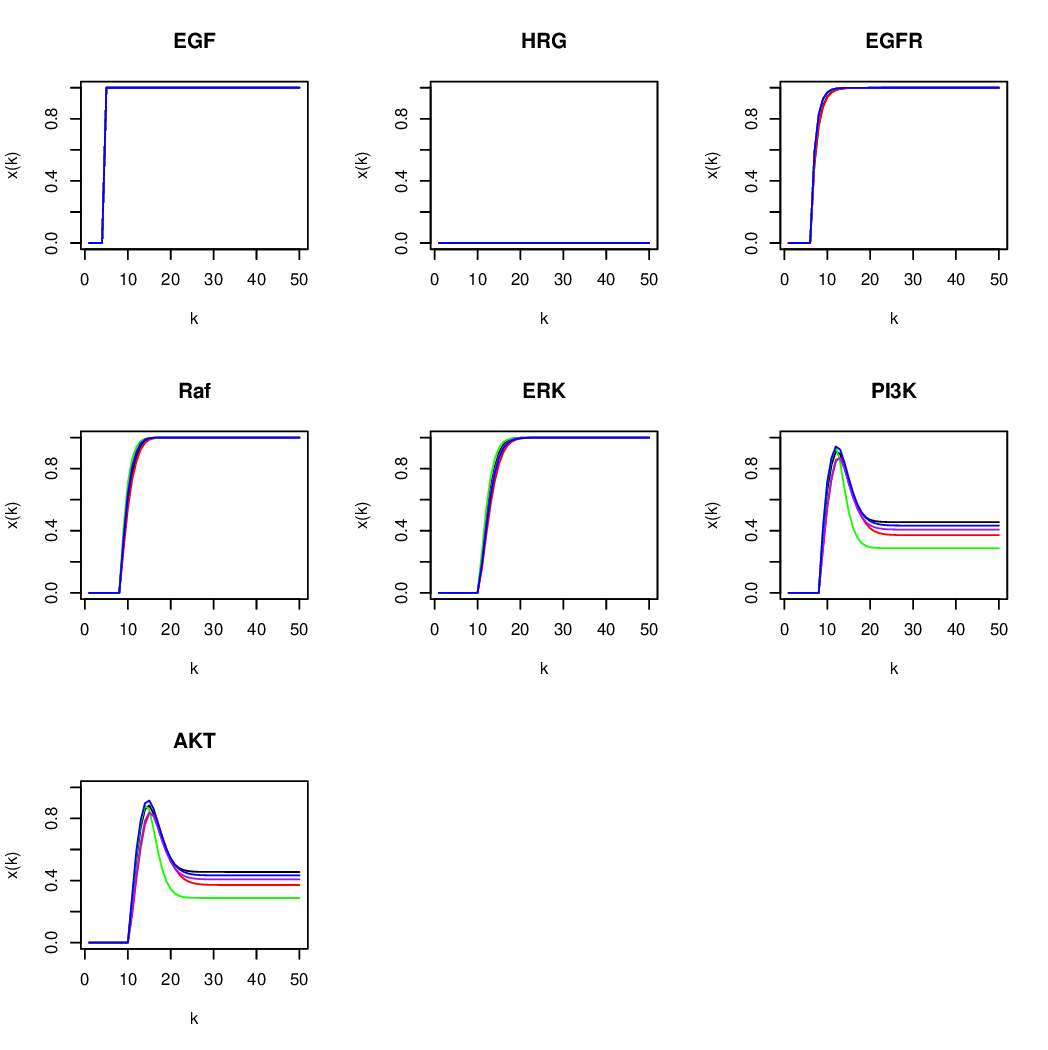}
\caption[Simulation 2]{\label{simulation2} Simulation 2: weakening the inhibitory effect of $ERK$ on $PI3K$.}
\end{center}
\end{figure*}

To implement this perturbation, the parameter values are as in simulation 1 except $w_{ERK,PI3K}$ which is set to $weaker$. With $w_{ERK,PI3K}=weaker$, the signal conveyed by the edge $(ERK,PI3K)$ is weakened throughout this simulation. The corresponding results are shown in \hyperref[simulation2]{\texttt{Figure \ref*{simulation2}}} page \pageref{simulation2}. As expected, weakening the edge $(ERK,PI3K)$ results in a weakened inhibition of $PI3K$ by $ERK$: $ERK$ does not totally inhibit $PI3K$.

\subsection{Simulation 3}

\begin{figure*}[!hbp]
\begin{center}
\includegraphics[width=1\textwidth]{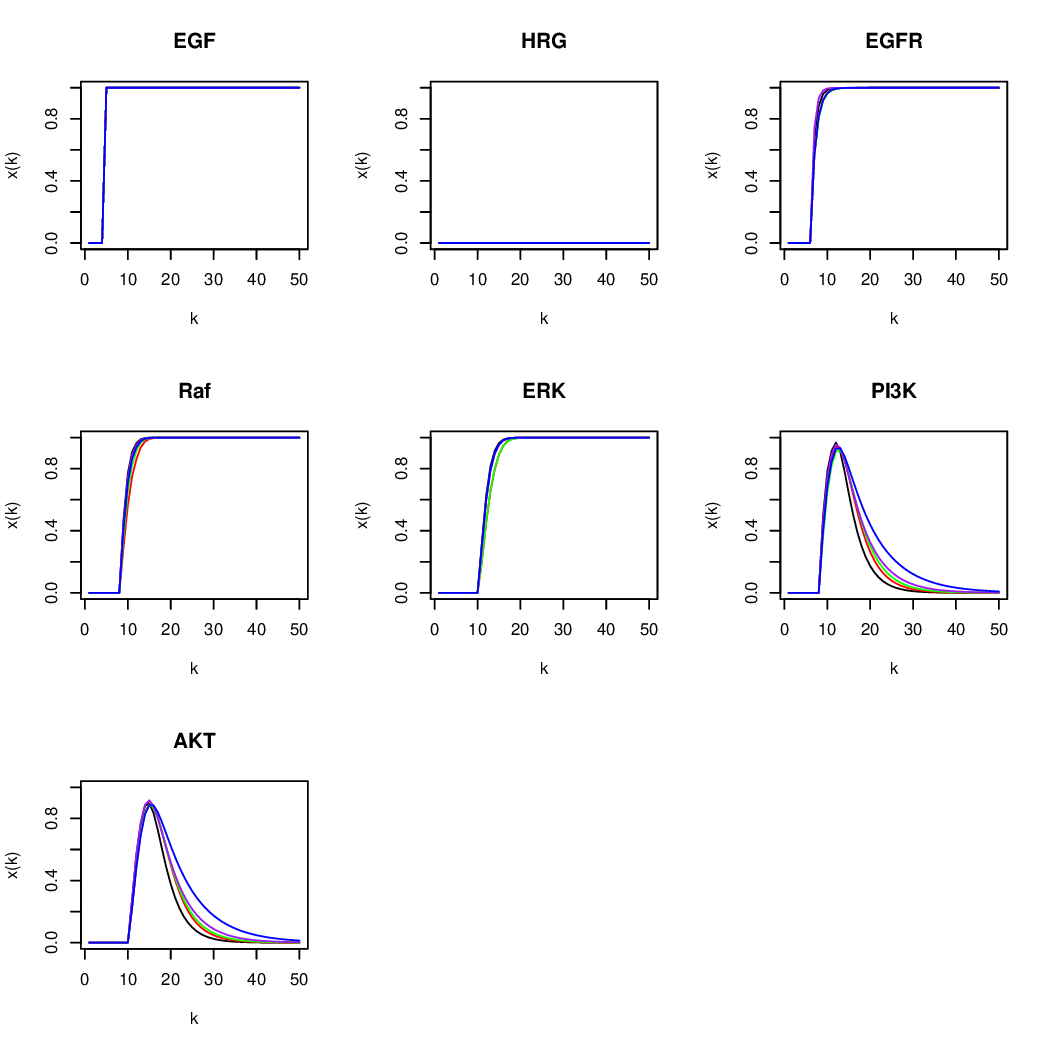}
\caption[Simulation 3]{\label{simulation3} Simulation 3: slowing the inhibitory effect of $ERK$ on $PI3K$.}
\end{center}
\end{figure*}

A perturbation is again applied on the edge $(ERK,PI3K)$. However, in this simulation the perturbation concerns its reactivity, namely $r_{ERK,PI3K}$ which is set to $slower$. The other parameter values are as in simulation 1. With $r_{ERK,PI3K}=slower$, the signal conveyed by the edge $(ERK,PI3K)$ is slowed throughout this simulation. The corresponding results are shown in \hyperref[simulation3]{\texttt{Figure \ref*{simulation3}}} page \pageref{simulation3}. As expected, slowing the edge $(ERK,PI3K)$ results in a slowed inhibition of $PI3K$ by $ERK$: although $ERK$ totally inhibits $PI3K$, it does it slower than in simulation 1 where $r_{ERK,PI3K}=fast$.

\subsection{Simulation 4}

In this simulation, no perturbations are applied and the parameter values are as in simulation 1. However, rather than totally activating $EGF$, it is set to $few$. Therefore, $f^{node}_{EGF}$ and $f^{node}_{HRG}$ become:

\begin{small}
\begin{IEEEeqnarray*}{r C l}
EGF(k+1)&=&
\begin{cases}
few&\text{if }k \geq k_{EGF}\\
none&\text{if }k<k_{EGF}
\end{cases}\\
HRG(k+1)&=&none
\end{IEEEeqnarray*}
\end{small}

\begin{figure*}[!hbp]
\begin{center}
\includegraphics[width=1\textwidth]{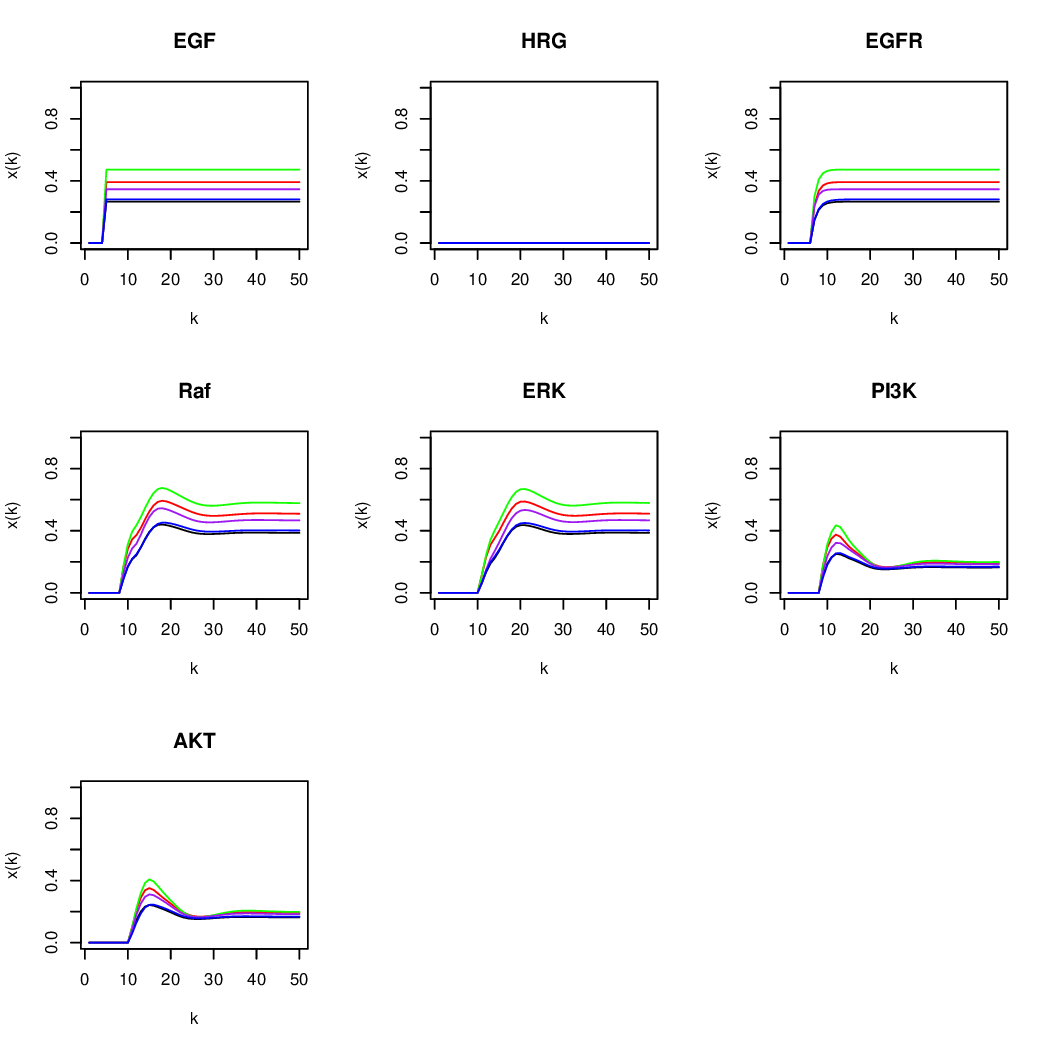}
\caption[Simulation 4]{\label{simulation4} Simulation 4: partial activation of $EGF$.}
\end{center}
\end{figure*}

The corresponding results are shown in \hyperref[simulation4]{\texttt{Figure \ref*{simulation4}}} page \pageref{simulation4}. As expected, the activation of $EGF$ is not total and so the same applies to the entire signaling cascade. For example, $PI3K$ is not totally activated since $EGFR$ is not totally activated too. Furthermore, $PI3K$ is not totally inhibited by $ERK$ because $ERK$ itself is not totally activated.

\subsection{Simulation 5}

\begin{figure*}[!hbp]
\begin{center}
\includegraphics[width=1\textwidth]{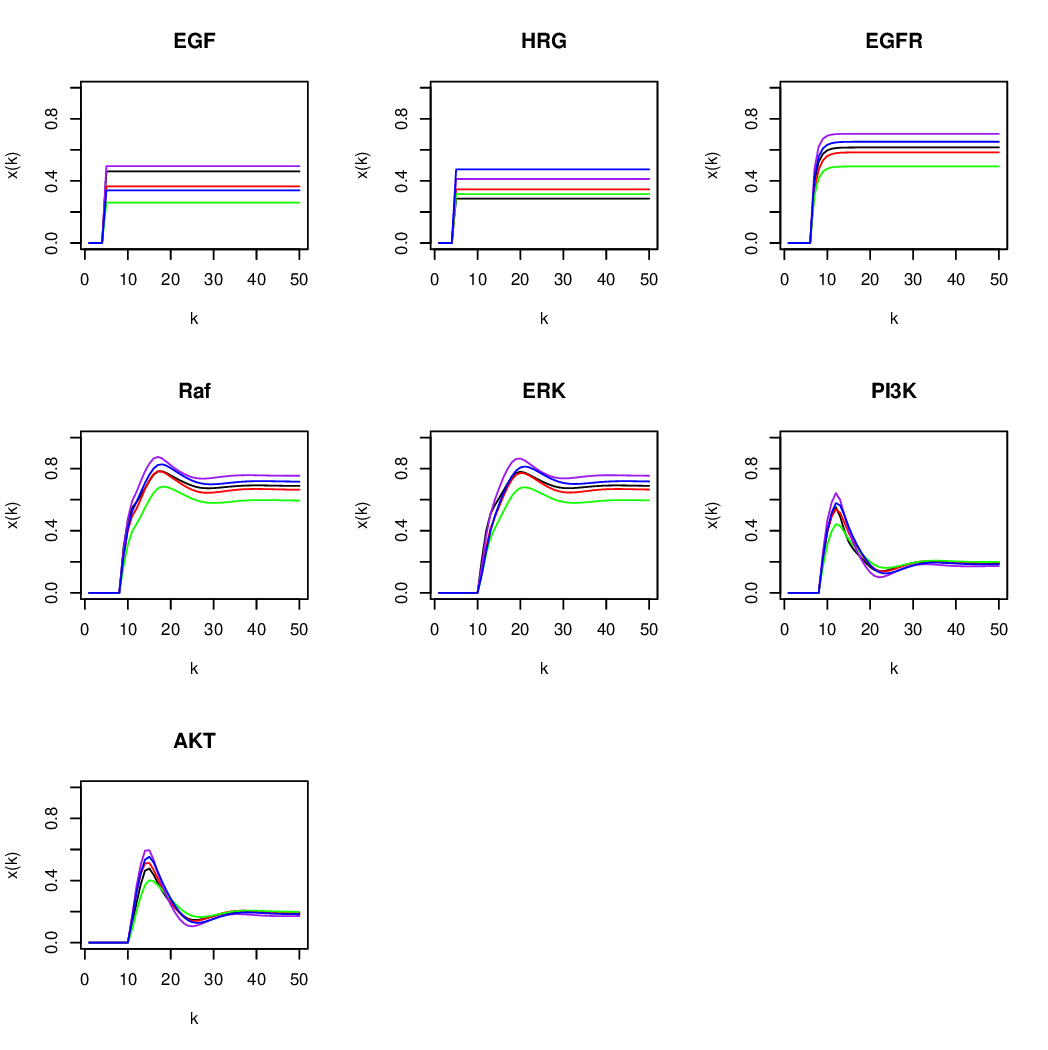}
\caption[Simulation 5]{\label{simulation5} Simulation 5: partial activation of $EGF$ and $HRG$.}
\end{center}
\end{figure*}

In this simulation, both $EGF$ and $HRG$ are set to $few$. Therefore, $f^{node}_{EGF}$ and $f^{node}_{HRG}$ become:

\begin{small}
\begin{IEEEeqnarray*}{r C l}
EGF(k+1)&=&
\begin{cases}
few&\text{if }k \geq k_{EGF}\\
none&\text{if }k<k_{EGF}
\end{cases}\\
HRG(k+1)&=&
\begin{cases}
few&\text{if }k \geq k_{HRG}\\
none&\text{if }k<k_{HRG}
\end{cases}
\end{IEEEeqnarray*}
\end{small}

\noindent with $k_{HRG}=k_{EGF}$, the other parameter values being as in simulation 1. The corresponding results are shown in \hyperref[simulation5]{\texttt{Figure \ref*{simulation5}}} page \pageref{simulation5}.

It points out that the effects of $EGF$ and $HRG$ on $EGFR$ are cumulative due to an $OR$ gate. Indeed, although both $EGF$ and $HRG$ are set to $few$, cumulating their effects on $EGFR$ makes the signaling cascade more active than in simulation 4 where only $EGF$ is set to $few$.

\section{Conclusion}
Owing to the use of the logical operators taken from fuzzy logic, the simulations performed with the example show that the proposed logic-based modeling formalism is able to produce continuous results while remaining qualitative. This allows qualitative variables to be more finely valued than with discrete approaches, such as Boolean networks, by taking into account all the possible levels. Moreover, thanks to the additional features \textit{edge reactivity} and \textit{edge weakening} attributed to each edge of the network, it is possible to tune in speed and strength the interactions taking place in the modeled biological system according to knowledge about them. This is expected to take into account that some interactions can be weaker or slower relative to other ones and therefore to be more realistic in their qualitative modeling.

These enhancements should enable to incorporate more knowledge, notably about biological processes, and to obtain more accurate results. In exchange, they require the parameters controlling how the signal flows in the edges to be valued. These parameters are intended to be qualitative, that is parameters whose the valuation is knowledge-based, by opposition to quantitative parameters whose the valuation is data-based. In other words, qualitative parameters translate qualitative information, an information which should be easier to obtain than the quantitative one. Indeed, quantitative models require their parameters to be valued by data obtained through experimental measurements. However, due to experimental limitations, such measurements can be challenging. Qualitative information is easier to obtain but at the cost of being qualitative, as its name indicates.

A little of stochasticity on the two additional features \textit{edge reactivity} and \textit{edge weakening} is also realized through the random selection of their value in specified intervals followed by replication and superposition of the produced results. This stochasticity, although very rudimentary, constitutes a line of improvement which should yield more realism because events taking place in biological systems are themselves subjected to stochasticity \cite{szekely2014stochastic,buiatti2013randomness,ullah2010stochastic,shahrezaei2008stochastic,kurakin2005stochastic}.

Another improvement could be to apply information theory \cite{shannon1948information} on the signal conveyed by the edges, as previously introduced for cell signaling \cite{mc2014information,brennan2012information,rhee2012application,waltermann2011information}. This improvement should enable to better model how the information flows in biological networks and particularly, starting from its sender, how the information is altered by noise before reaching its receiver. Such information alterations could have significant consequences on the functionalities of biological networks, such as inappropriate responses to extracellular signals.

Altogether, starting from Boolean networks and still founded on their basic principles, this work is expected to bring a fine qualitative quantification of biological network behaviors. It should be noted that a qualitative quantification remains qualitative and should not be confused with a true quantification which involves experimental measurements, values and units \cite{benenson2002measurements}. The qualitative quantification proposed in the present work has the goal of bringing enhancements in the ability of qualitative models to simulate the behavior of biological networks. One of the main goals, and advantages, of qualitative modeling remains to propose an alternative to, but not a replacement of, quantitative approaches when the frequently encountered scarcity in quantitative information makes the work unreasonably or unnecessarily difficult.

It is also possible to use qualitative and quantitative approaches in combination. For example, qualitative modeling can be used to explore global properties and then quantitative modeling can be used to focus on particular events. Knowing the difficulty of quantitative modeling in systems biology, this two-steps approach could make modeling more efficient by highlighting where to deploy quantitative approaches. Qualitative and quantitative approaches can also be merged into hybrid models \cite{khan2014hybrid,berestovsky2013modeling,samaga2013modeling,glass1973logical} which attempt to exploit the advantages of these two formalisms in one. Hybrid models, or semi-quantitative models, can be good compromises between the convenience of qualitative modeling and the accuracy of quantitative modeling.

Finally, continuous dynamical systems are frequently modeled by differential equations for which advanced solvers are available, such as LSODE (the Livermore Solver for Ordinary Differential Equations) \cite{hindmarsh1983odepack}. The present work introduces continuous dynamical systems made of logical equations. However, mathematically speaking, it is likely that these continuous logical equations are differential equations thought and built in a different way. Consequently, it would be possible to mathematically express them as differential equations and then to use available computational tools aimed at simulating and analyzing continuous dynamical systems.

\bibliographystyle{unsrt}
\bibliography{enhancing_boolean_networks}

\end{document}